\documentclass[11pt]{article}

\usepackage[margin=1in]{geometry}

\usepackage{graphicx} 

\usepackage[square,numbers]{natbib}

\usepackage{amsmath}
\usepackage{amsfonts}

\usepackage{booktabs}
\usepackage{url}

\usepackage[hang,small,bf]{caption}
\usepackage[subrefformat=parens]{subcaption}
\usepackage{hyperref}

\begin{document}

\begin{center}
    {\Large\bfseries
        Generative climate downscaling enables high-resolution compound risk assessment by preserving multivariate dependencies
    }

    \vspace{1em}

    Takuro Kutsuna$^{1}$, Noriko N. Ishizaki$^{2}$, Norihiro Oyama$^{1}$, Hiroaki Yoshida$^{1}$

    \vspace{0.5em}

    {\small
        $^{1}$Toyota Central R\&D Labs., Inc. \\
        $^{2}$National Institute for Environmental Studies
    }

\end{center}

\vspace{0em}

\section*{Abstract}
Physics-based climate projections using general circulation models are essential for assessing future risks, but their coarse resolution limits regional decision-making. Statistical downscaling can efficiently add detail, yet many methods treat variables independently, degrading inter-variable relationships that govern compound hazards such as heat stress, drought, and wildfire.
Here we show that a diffusion-based multivariate generative framework, combined with bias correction, recovers degraded inter-variable correlations even under a 50$\times$ increase in linear resolution.
When applied to five meteorological variables over Japan, the framework reduces inter-variable correlation errors by more than fourfold relative to existing baselines while improving both univariate and spatial accuracy, leading to more accurate detection of severe drought.
These results demonstrate that multivariate generative downscaling improves the reliability of compound risk assessment under large resolution gaps.

\section{Introduction} \label{sec:introduction}
Physics-based climate system simulations are widely used to project climate change over coming decades. Because Earth’s climate is governed by global circulation, long-term projections rely on whole-Earth simulations using General Circulation Models (GCMs) \citep{CMIP5,CMIP6}. However, computational constraints limit GCMs to relatively coarse spatial resolution, particularly for multi-decadal to centennial simulations; for example, many models participating in the sixth phase of the Coupled Model Intercomparison Project (CMIP6) \citep{CMIP6} typically have horizontal resolutions on the order of 100 km, which is insufficient to provide the fine-scale climate information needed for regional impact studies and adaptation planning.
To bridge this resolution gap, downscaling (DS) methods generate higher-resolution climate information from GCM outputs. DS methods are broadly classified into dynamical and statistical approaches.
Dynamical DS employs high-resolution regional climate models driven by GCM boundary conditions \citep{giorgi1989climatological,deque2005global,kawase2008downscaling}, providing strong physical consistency at high computational cost, whereas statistical DS infers relationships between local-scale predictands and large-scale predictors derived from GCM outputs or reanalysis \citep{wilby2004guidelines,von1993downscaling,piani2010statistical,iizumi2011evaluation}.

Climate projections play an important role in societal risk assessments, motivating careful characterization of environmental hazards, many of which are compound in nature and emerge when multiple meteorological variables jointly meet specific conditions.
For example, human heat stress is governed not only by air temperature but also by humidity, radiation, and wind speed, as reflected in integrated indices such as the Wet Bulb Globe Temperature (WBGT) \citep{iso7243}.
Similarly, drought and wildfire risks are strongly influenced by the combined effects of precipitation and atmospheric conditions, and are commonly evaluated using multivariate indices such as the Standardized Precipitation Evapotranspiration Index (SPEI) \citep{vicente2010multiscalar} and the Fire Weather Index (FWI) \citep{wagner1987development}.
These examples illustrate that many high-impact climate risks are inherently linked to the co-variability and correlation structure among meteorological variables.

While computationally efficient and scalable, many statistical DS methods---such as regression-based approaches \citep{wilby2002sdsm,chen2010downscaling,davy2010statistical} and distribution-based bias correction methods such as quantile mapping (QM) \citep{wood2004hydrologic,cannon2015bias}---are applied independently for each variable and each location, making it difficult to reproduce local spatial correlations and multivariate dependencies. Although multivariate extensions have been proposed \citep{cannon2018multivariate,bardossy2012multiscale}, they often become unstable as dimensionality increases with the number of variables and grid points \citep{esd-11-537-2020}.
Recent advances in deep neural networks (DNNs) have motivated DNN-based DS methods \citep{sun2024deep}, particularly super-resolution approaches that improve the reproduction of spatial structures \citep{oyama2023deep}.
However, most existing studies focus on a single variable, and even multivariate extensions typically involve only a limited number of variables \citep{sun2024deep}.
More recently, diffusion models and their variants have attracted increasing attention as a promising framework for statistical DS \citep{leinonen2023latent,ling2024diffusion,mardani2025residual,hess2025fast,lopez2025dynamical}.
Most existing studies emphasize reconstruction fidelity at the variable level, whereas explicit assessments of cross-variable dependencies remain scarce.
A recent hybrid framework, in which coarse-resolution DS is first performed using dynamical DS and then refined using diffusion-based statistical DS, has shown promise for improving the reproducibility of inter-variable correlations in downscaled fields \citep{lopez2025dynamical}.
However, the overall DS factor in that framework remains modest, with the diffusion-based component providing only a 5$\times$ refinement, and the inclusion of dynamical DS inevitably sacrifices computational efficiency.
Thus, it remains unclear whether downscaling at larger, more practically relevant scale factors can be achieved by purely statistical DS while reliably preserving multivariate structure; systematic evaluations of this issue also remain insufficiently established.

\begin{figure}[htb]
    \centering
    \includegraphics[width=1.0\linewidth,clip]{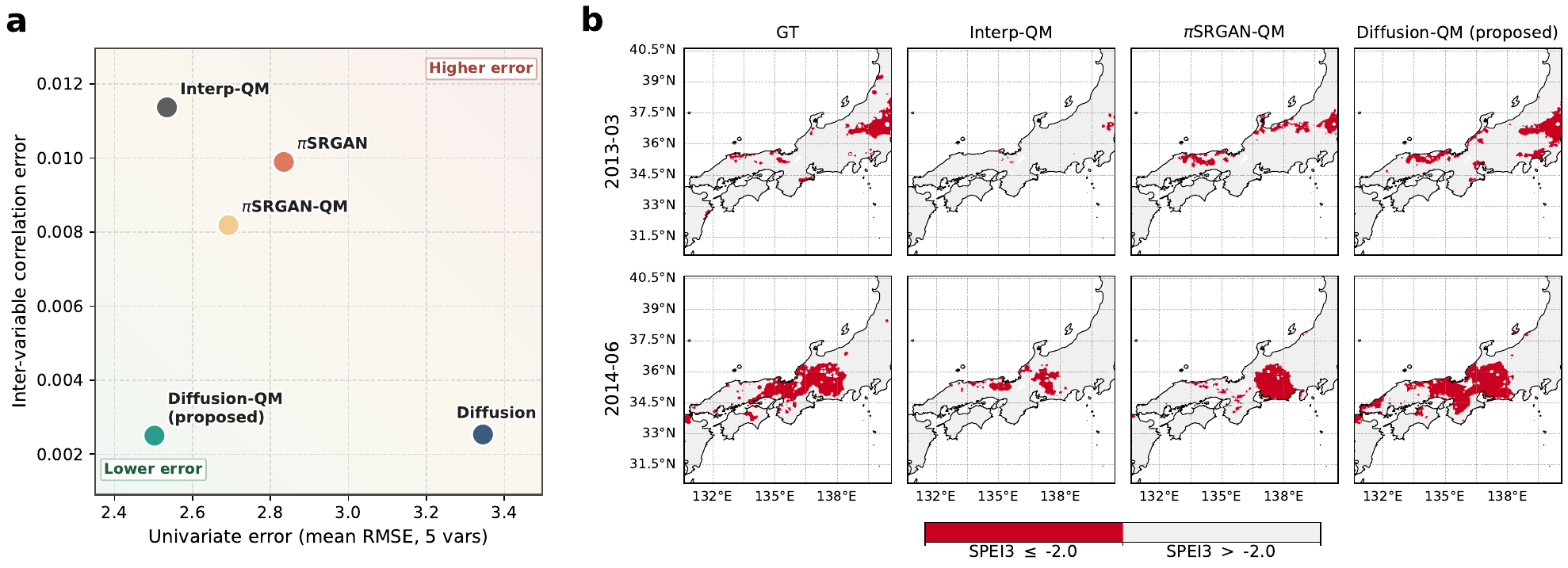}
    \caption{\textbf{a} Comparison of DS methods in terms of univariate prediction error (RMSE averaged over the five target variables) and inter-variable correlation error. \textbf{b} Maps showing areas under severe drought conditions, defined by $\mathrm{SPEI3} \leq -2.0$, for two representative months from the test period (2013-03 and 2014-06), comparing GT with estimates derived from the downscaled outputs of Interp-QM, $\pi$SRGAN-QM, and Diffusion-QM.}
    \label{fig:error_and_spei_threshold_201303_201406_all_m2p0}
\end{figure}

In this study, we address this limitation by demonstrating that diffusion-based multivariate statistical DS can recover inter-variable correlation structures degraded in low-resolution inputs, even under a large 50$\times$ increase in linear resolution (from 1.25$^\circ$ to 0.025$^\circ$ latitude--longitude grid spacing in our setup).
We downscale five meteorological variables over Japan using coarse-resolution reanalysis predictors \citep{kobayashi2015jra} as a controlled testbed. High-resolution observations \citep{OHNO2016J-16-028} are used as reference data for evaluating univariate accuracy and multivariate dependencies.
As summarized in Fig.~\ref{fig:error_and_spei_threshold_201303_201406_all_m2p0}, combining diffusion-based multivariate generation with QM substantially reduces inter-variable correlation errors while maintaining low univariate prediction error, and improves the identification of severe drought events.
These results demonstrate that generative multivariate downscaling improves the reliability of high-resolution compound risk assessment under large resolution gaps.

\section{Results} \label{sec:results}

\subsection{Overall DS accuracy across meteorological variables}

We first assess overall DS accuracy using the root mean square error (RMSE) for each target variable---daily mean temperature (Tmean), daily maximum temperature (Tmax), daily minimum temperature (Tmin), precipitation (Precip), and global solar radiation (GSR)---averaged over the test period (Table~\ref{tab:res_rmse}).
Applying QM to diffusion-based outputs reduces RMSE for all variables except Precip, and Diffusion-QM achieves the lowest RMSE for most targets.
In contrast, the diffusion-only model exhibits larger errors for some variables, notably Tmax and GSR.
Definitions and units of the target variables are summarized in Supplementary Table~\ref{tab:s_hr_variables}.

\begin{table}[htb]
    \centering
    \caption{Prediction errors (RMSE) of each DS method for each target variable.}
    \label{tab:res_rmse}
    \begin{tabular}{lccccc}
        \toprule
        Method & Tmean & Tmax  & Tmin  & Precip & GSR   \\
        \midrule
        Interp-QM
               & 0.952 & 1.513 & 1.569 & 5.944  & 2.696 \\
        $\pi$SRGAN
               & 1.598 & 2.009 & 1.883 & 5.987  & 2.696 \\
        $\pi$SRGAN-QM
               & 1.221 & 1.743 & 1.639 & 6.226  & 2.632 \\
        Diffusion
               & 1.810 & 3.427 & 1.600 & 5.950  & 3.938 \\
        Diffusion-QM
               & 0.950 & 1.390 & 1.335 & 6.224  & 2.614 \\
        \bottomrule
    \end{tabular}
\end{table}

\begin{figure}[tbp]
    \centering
    \includegraphics[width=1.0\linewidth,clip]{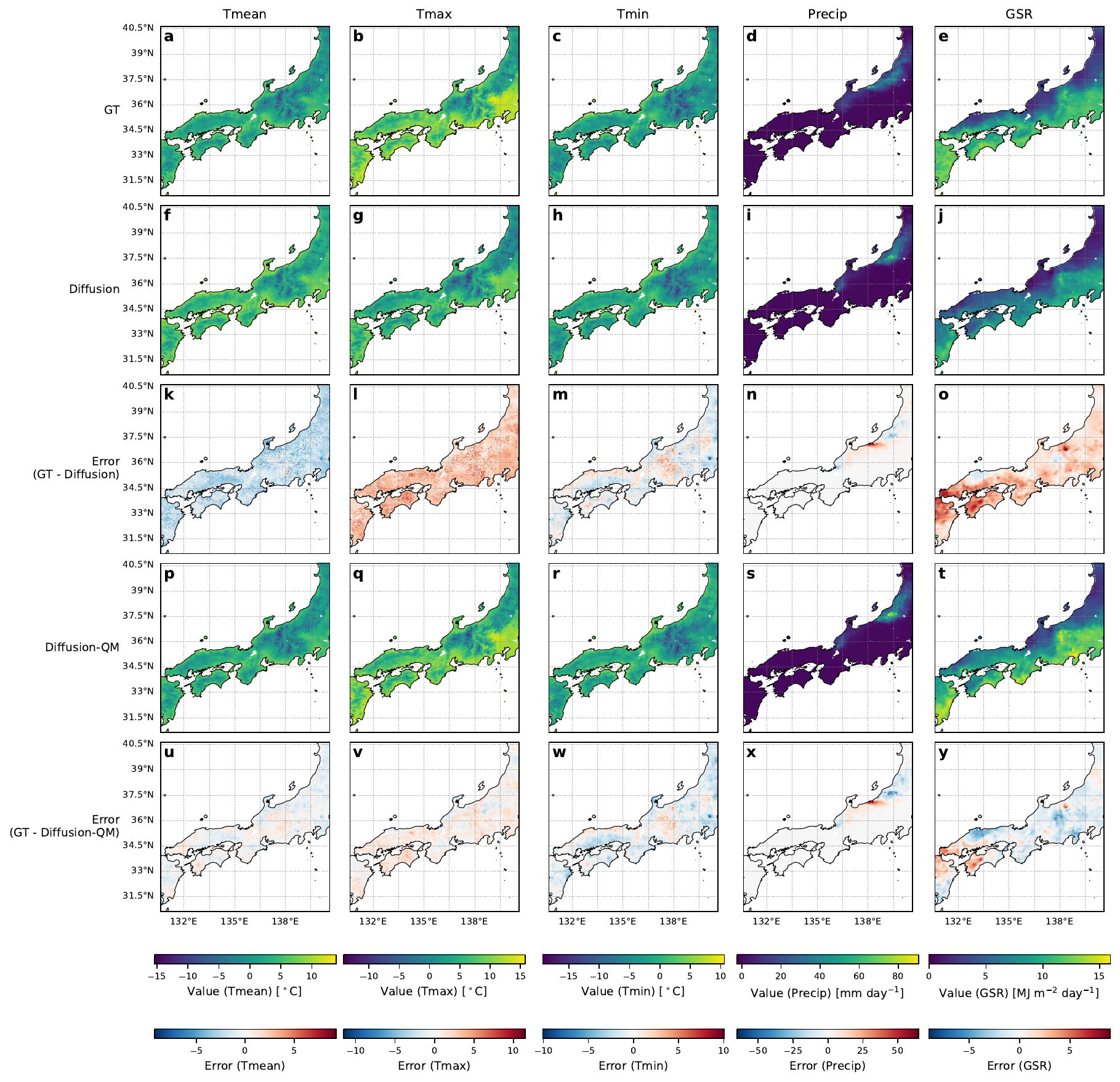}
    \caption{Downscaled fields and error maps for a representative day from the test period (2023-01-01). Columns correspond to Tmean, Tmax, Tmin, Precip, and GSR. Units are shown in the color bars. From top to bottom: GT, Diffusion, signed error (GT $-$ Diffusion), Diffusion-QM, and signed error (GT $-$ Diffusion-QM).}
    \label{fig:results_ds_qm_errors_2023-01-01}
\end{figure}

While Table~\ref{tab:res_rmse} summarizes mean performance over the test period,
it does not capture the spatial distribution of errors on individual days.
We therefore visualize a representative test case (2023-01-01) and compare the
downscaled fields produced by Diffusion and Diffusion-QM with the ground truth
(GT), together with the corresponding error maps
(Fig.~\ref{fig:results_ds_qm_errors_2023-01-01}).
For Tmean and Tmax, the Diffusion results exhibit spatially coherent biases
(Fig.~\ref{fig:results_ds_qm_errors_2023-01-01}k,l), which are substantially
reduced after applying QM
(Fig.~\ref{fig:results_ds_qm_errors_2023-01-01}u,v).
The corresponding low-resolution reanalysis fields used for conditioning are
shown in Supplementary Fig.~\ref{fig:s_lr_images_2023-01-01}.

\subsection{Inter-variable correlation structures}

Beyond sample-wise univariate errors, we assess how well the downscaled meteorological fields preserve inter-variable correlation structures.
As summarized in Table~\ref{tab:ds_intervar_corr}, diffusion-based approaches achieve the smallest errors across all seasons and in the overall average.
These errors are computed between the DS outputs and the GT for each season (DJF: Dec--Feb, MAM: Mar--May, JJA: Jun--Aug, and SON: Sep--Nov), and the seasonal mean (AVG) is also reported.

\begin{table}[t]
    \centering
    \caption{Comparison of inter-variable correlation errors across DS methods. Errors are evaluated separately for each season, and AVG denotes the seasonal mean of the corresponding errors.}
    \label{tab:ds_intervar_corr}
    \begin{tabular}{lccccc}
        \toprule
        Method        & DJF    & MAM    & JJA    & SON    & AVG    \\
        \midrule
        Interp-QM     & 0.0254 & 0.0101 & 0.0059 & 0.0041 & 0.0114 \\
        $\pi$SRGAN    & 0.0168 & 0.0059 & 0.0126 & 0.0042 & 0.0099 \\
        $\pi$SRGAN-QM & 0.0132 & 0.0056 & 0.0105 & 0.0035 & 0.0082 \\
        Diffusion     & 0.0036 & 0.0013 & 0.0035 & 0.0018 & 0.0025 \\
        Diffusion-QM  & 0.0029 & 0.0018 & 0.0039 & 0.0014 & 0.0025 \\
        \bottomrule
    \end{tabular}
\end{table}

Diffusion and Diffusion-QM yield nearly identical errors, indicating that the
subsequent application of QM does not substantially alter inter-variable
correlation structures in our experiments.
This behavior contrasts with the RMSE results (Table~\ref{tab:res_rmse}), where
larger differences are observed between Diffusion and Diffusion-QM.

\subsection{Detailed analysis of inter-variable relationships}
\begin{figure}[tbp]
    \centering
    \includegraphics[width=1.0\linewidth,clip]{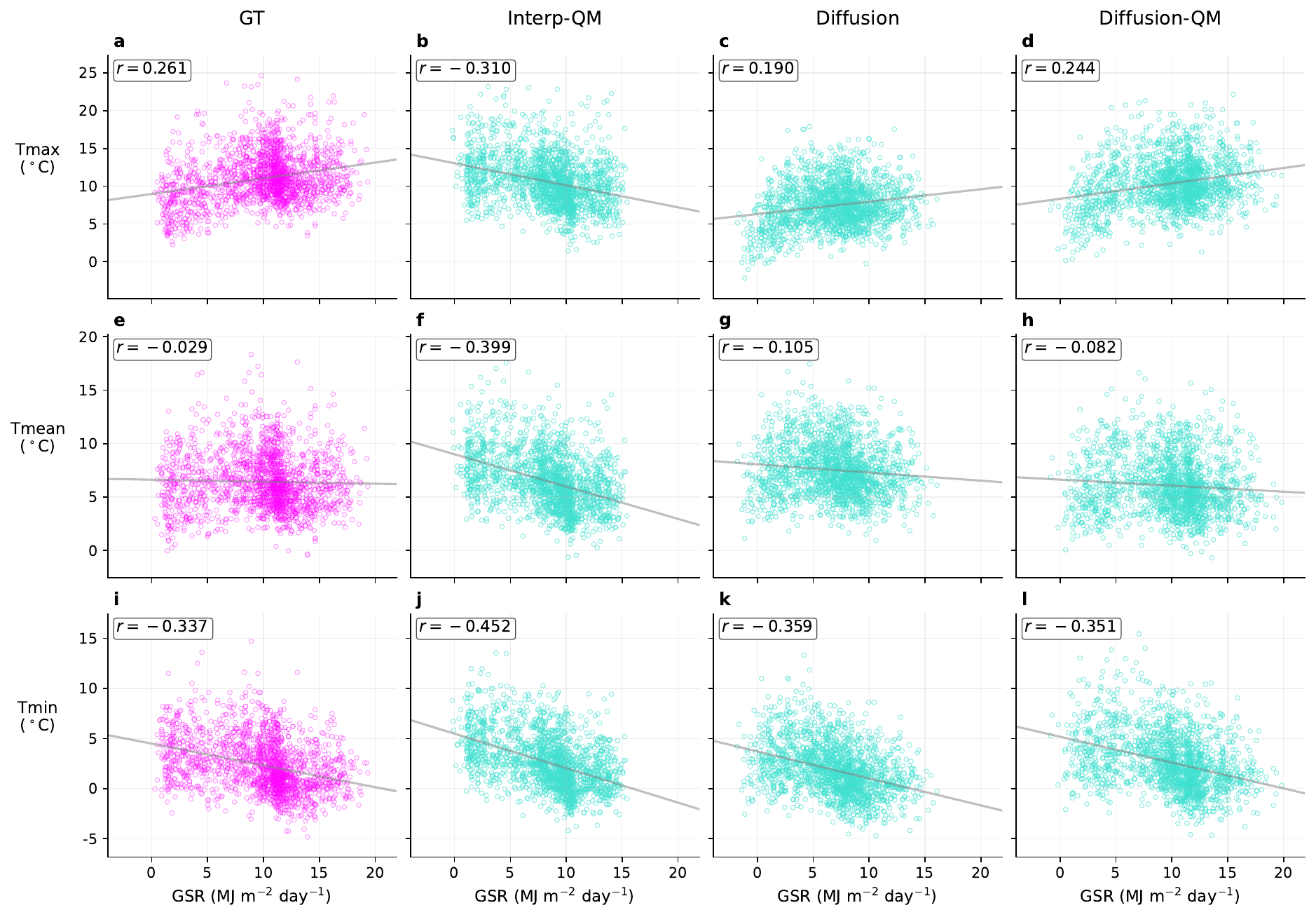}
    \caption{Scatter plots showing the relationship between GSR (horizontal axis) and air temperature (vertical axis) in Tokyo during winter (DJF) over the test period. From top to bottom, the panels show relationships with Tmax, Tmean, and Tmin, respectively. Each point represents an individual day in the test period. In each row, the leftmost panel shows the relationship in the GT high-resolution observations, followed by the downscaled results obtained using Interp-QM, Diffusion, and Diffusion-QM. The correlation coefficient between the two variables is indicated in the upper-left corner of each panel, and the solid line denotes the linear regression fit.}
    \label{fig:results_scatter_corr_Tokyo_gsr_temps}
\end{figure}

To provide a more interpretable view of the inter-variable dependence captured by each method, we examine a representative relationship at a selected location. Fig.~\ref{fig:results_scatter_corr_Tokyo_gsr_temps} shows scatter plots of daily GSR versus air temperature in Tokyo during winter (DJF) over the test period. From top to bottom, the panels correspond to Tmax, Tmean, and Tmin; from left to right, they compare GT with Interp-QM, Diffusion, and Diffusion-QM. The correlation coefficient and a linear regression fit are shown in each panel.

In GT, GSR exhibits a positive correlation with Tmax, an approximately zero correlation with Tmean, and a negative correlation with Tmin, highlighting distinct dependence patterns across temperature statistics (Fig.~\ref{fig:results_scatter_corr_Tokyo_gsr_temps}a,e,i). Interp-QM fails to capture these sign-dependent relationships, showing negative correlations between GSR and all three temperature variables (Fig.~\ref{fig:results_scatter_corr_Tokyo_gsr_temps}b,f,j). In contrast, Diffusion-QM shows closer agreement with GT across the three panels, recovering the positive/near-zero/negative correlation pattern for Tmax, Tmean, and Tmin (Fig.~\ref{fig:results_scatter_corr_Tokyo_gsr_temps}d,h,l). The Diffusion results capture the qualitative correlation tendency but exhibit systematic offsets along both axes relative to GT (Fig.~\ref{fig:results_scatter_corr_Tokyo_gsr_temps}c,g,k), which are substantially reduced after applying QM.

\subsection{Precipitation distribution and spatial correlation}

As an additional macro-scale assessment, we evaluate precipitation using two metrics, following \cite{oyama2023deep}, that quantify mismatches in marginal distributions (distribution KLD) and discrepancies in spatial correlation structures (spatial correlation error).

As summarized in Table~\ref{tab:ds_precip_macro}, the diffusion-based approaches yield markedly smaller spatial correlation errors than Interp-QM and the $\pi$SRGAN-based baselines; in particular, Diffusion-QM reduces this error by approximately eightfold relative to Interp-QM (0.0377 to 0.0048). For the distribution metric, the lowest KLD is achieved by Diffusion-QM.

\begin{table}[htbp]
    \centering
    \caption{Comparison of precipitation-related macro-scale metrics across DS methods.}
    \label{tab:ds_precip_macro}
    \begin{tabular}{lcc}
        \toprule
        Method        & Distribution KLD & Spatial correlation error \\
        \midrule
        Interp-QM     & 0.0056           & 0.0377                    \\
        $\pi$SRGAN    & 0.0053           & 0.0163                    \\
        $\pi$SRGAN-QM & 0.0034           & 0.0139                    \\
        Diffusion     & 0.0034           & 0.0051                    \\
        Diffusion-QM  & 0.0023           & 0.0047                    \\
        \bottomrule
    \end{tabular}
\end{table}

\subsection{Drought-risk assessment based on SPEI3}

To assess whether improved multivariate consistency translates into improved compound-risk diagnostics, we evaluate drought detection based on the 3-month SPEI (SPEI3). We binarize SPEI3 using three thresholds ($\leq -1.0$, $\leq -1.5$, and $\leq -2.0$) and compute the intersection over union (IoU) between the GT and the DS-based estimates over the test period. Figure~\ref{fig:error_and_spei_threshold_201303_201406_all_m2p0}b shows representative maps for the $\leq -2.0$ threshold.

As summarized in Table~\ref{tab:drought_iou}, Diffusion-QM achieves the highest IoU at all three thresholds, followed by $\pi$SRGAN-QM and Interp-QM. The performance gap becomes larger at more severe drought thresholds; for example, at the $\leq -2.0$ threshold, the IoU increases from 0.262 for Interp-QM to 0.324 for Diffusion-QM.

\begin{table}[htb]
    \centering
    \caption{Comparison of drought detection accuracy based on binarized SPEI3. Column headers indicate the SPEI3 thresholds used for binarization. IoU is computed over the test period between GT and DS-based estimates, and higher values indicate better agreement.}
    \label{tab:drought_iou}
    \begin{tabular}{lccc}
        \toprule
        Method        & \multicolumn{3}{c}{SPEI3 binarization threshold}                             \\
        \cmidrule(lr){2-4}
                      & $\leq -1.0$                                      & $\leq -1.5$ & $\leq -2.0$ \\
        \midrule
        Interp-QM     & 0.444                                            & 0.335       & 0.262       \\
        $\pi$SRGAN    & 0.396                                            & 0.300       & 0.232       \\
        $\pi$SRGAN-QM & 0.479                                            & 0.380       & 0.310       \\
        Diffusion     & 0.403                                            & 0.281       & 0.203       \\
        Diffusion-QM  & 0.498                                            & 0.392       & 0.324       \\
        \bottomrule
    \end{tabular}
\end{table}

Supplementary Fig.~\ref{fig:s_fig_monthly_spei_drought_qq} further shows monthly SPEI3 statistics and quantile--quantile comparisons, providing a more detailed view of the agreement between GT and the Diffusion-QM estimates over the test period.

\subsection{Extreme-event case study}

\begin{figure}[tbp]
    \centering
    \includegraphics[width=1.0\linewidth,clip]{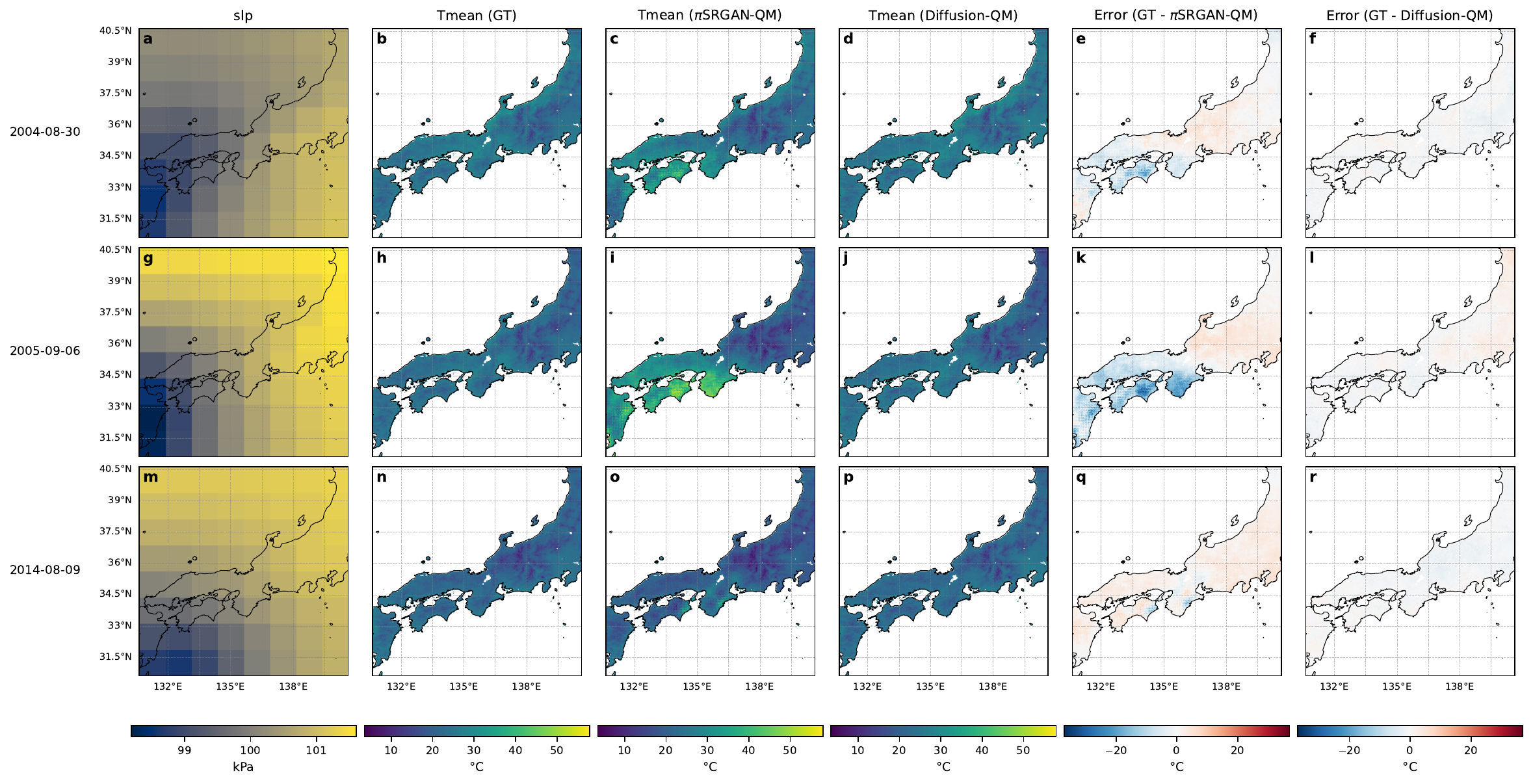}
    \caption{Case-study comparison of downscaled Tmean fields during typhoon-related events in the test period. Rows show three representative dates (2004-08-30, 2005-09-06, and 2014-08-09). Columns show slp (an LR reanalysis predictor, upsampled from $8\times8$ to $400\times400$), GT Tmean, $\pi$SRGAN-QM Tmean, Diffusion-QM Tmean, and the corresponding error maps.}
    \label{fig:compare_typhoon_ds}
\end{figure}

To assess the robustness of the proposed method under extreme meteorological conditions, we examine DS performance using three typhoon-related days from the test period (2004-08-30, 2005-09-06, and 2014-08-09).
Fig.~\ref{fig:compare_typhoon_ds} presents a comparison of the Tmean fields produced by $\pi$SRGAN-QM and Diffusion-QM, together with the corresponding signed error maps, for each case.

In all three events, the sea-level pressure (slp) fields used for conditioning indicate an approaching low-pressure system characteristic of typhoon conditions.
Under these extreme settings, $\pi$SRGAN-QM exhibits substantially larger spatial Tmean errors than Diffusion-QM, whereas Diffusion-QM maintains more physically consistent temperature fields with smaller deviations from GT.

\section{Discussion} \label{sec:discussion}

The large RMSE of the Diffusion outputs (Table~\ref{tab:res_rmse}) and the spatially coherent offsets in the Diffusion results (Fig.~\ref{fig:results_ds_qm_errors_2023-01-01}k,l) suggest that diffusion-only DS can retain systematic distributional biases.
Such behavior is consistent with known challenges of diffusion-based generation, including imperfect adherence to conditioning and a tendency to under-generate globally extreme patterns \citep{dhariwal2021diffusion,ho2021classifier,offset_noise,lin2024common}.
Several methods have been proposed to mitigate these issues in diffusion-based general-image generation, and applying similar techniques to meteorological DS may help alleviate these artifacts.
However, many such techniques introduce additional hyperparameters that require careful tuning.
Moreover, as shown in Supplementary Table~\ref{tab:s_val_rmse_cfg_offset}, the effects of these hyperparameters can differ substantially across variables, making it difficult in practice to identify a single setting that is simultaneously optimal for all target variables---particularly in our setting, where the proposed method performs DS for multiple variables using a single unified model.
For DS applications, reproducing statistical properties is typically more essential than improving perceptual or visual quality, and the climate community has accumulated extensive practical knowledge on bias correction for climate-model outputs.
Our two-stage design explicitly leverages the strengths of each component: diffusion modeling captures high-dimensional spatial and inter-variable correlation structure, while QM adjusts marginal distributions at each grid point using well-established bias-correction practice.
The inter-variable correlation errors are nearly unchanged after applying QM (Table~\ref{tab:ds_intervar_corr}), indicating that QM has little effect on these dependence diagnostics, whereas the RMSE improvements (Table~\ref{tab:res_rmse}) suggest that QM primarily corrects systematic biases and marginal distributional mismatches.
Although QM is applied independently to each variable and location, it largely preserves the rank ordering over time, which may help explain why inter-variable correlation errors remain similar for Diffusion and Diffusion-QM.

The Tokyo case study demonstrates that a learned conditional generative mapping can recover sign-dependent relationships between GSR and temperature (Tmax, Tmean, and Tmin) that are not reproduced by Interp-QM (Fig.~\ref{fig:results_scatter_corr_Tokyo_gsr_temps}).
A plausible explanation for the negative correlation between GSR and Tmax obtained with Interp-QM (Fig.~\ref{fig:results_scatter_corr_Tokyo_gsr_temps}b) is that the coarse resolution of the driving reanalysis fields can conflate heterogeneous surface types (e.g., land--sea mixtures within a single grid cell), thereby distorting the underlying physical relationships.
In general, ocean surfaces have a larger heat capacity than land and can exhibit a different temperature response to variations in incoming solar radiation; such mixed-surface effects may yield biased GSR--temperature dependencies that are subsequently propagated by the simple spatial interpolation used in Interp-QM.
In addition, biases in the representation of the planetary boundary layer in the driving reanalysis may affect near-surface temperature responses and the coupling between surface radiation and temperature.
In contrast, the diffusion model learns systematic, data-driven corrections from historical paired low- and high-resolution datasets, enabling the recovery of more physically consistent relationships at fine scales.

The improved performance of Diffusion-QM in the SPEI3-based drought assessment (Table~\ref{tab:drought_iou}) is physically plausible given how potential evapotranspiration (PET) is computed in the derivation of SPEI.
SPEI is computed from the climatic water balance, defined as precipitation minus PET. Accurate estimation of PET is therefore critical, as it directly controls the magnitude of the water balance and, consequently, the resulting SPEI values, particularly under dry conditions.
In our analysis, PET is estimated following \citep{hargreaves1985reference}, and depends on the diurnal temperature range. Consequently, errors in the joint distribution of Tmax and Tmin can propagate into PET estimates.
If a DS method fails to preserve the physical relationship between Tmax and Tmin, the resulting bias in diurnal temperature range can propagate to PET and subsequently to SPEI.
From this perspective, the improved preservation of inter-variable relationships by Diffusion-QM (Table~\ref{tab:ds_intervar_corr}; Fig.~\ref{fig:results_scatter_corr_Tokyo_gsr_temps}) is consistent with its enhanced drought-detection skill based on SPEI3.
At the same time, the diffusion-only results yield substantially lower IoU values than Diffusion-QM and are even worse than Interp-QM at all SPEI3 thresholds (Table~\ref{tab:drought_iou}), despite preserving inter-variable correlation errors at a level comparable to Diffusion-QM (Table~\ref{tab:ds_intervar_corr}).
This contrast suggests that preserving inter-variable relationships alone is not sufficient for accurate SPEI estimation when the univariate prediction errors of the individual variables remain large, as reflected in the higher RMSE of the diffusion-only outputs (Table~\ref{tab:res_rmse}).
Taken together, these results indicate that accurate SPEI estimation requires both preservation of physically meaningful inter-variable dependence and adequate univariate accuracy.

The typhoon case study (Fig.~\ref{fig:compare_typhoon_ds}) highlights another practical difference between DS approaches under rare conditioning states.
On the selected typhoon-related low-pressure days, $\pi$SRGAN-QM produces pronounced, spatially coherent Tmean errors, whereas Diffusion-QM remains much closer to the GT.
Because both methods apply the same QM post-processing, this contrast reflects differences in the underlying generative mapping.
One possible explanation is that a feed-forward GAN-based model effectively learns a direct nonlinear regression from LR predictors to HR targets; for low-frequency regimes such as typhoon-related extremes, the mapping may be weakly constrained by training samples and can yield unstable extrapolations.
By generating outputs through many incremental denoising steps conditioned on the LR inputs, diffusion sampling may provide a more stable mechanism for producing realistic fine-scale structure under such extreme conditions.

Beyond statistical performance, practical tractability is an important consideration for high-resolution DS.
Directly modeling high-resolution outputs with diffusion models is computationally challenging due to increasing GPU memory demands \citep{rombach2022high} and the difficulty of noise prediction at high resolution \citep{li2025jit}.
Following successful practice in modern image generation, we perform diffusion modeling in a low-dimensional latent representation defined by a pretrained autoencoder \citep{rombach2022high}.
Importantly, rather than training a task-specific autoencoder, we reuse an off-the-shelf model developed for general image generation, substantially reducing training cost while allowing our framework to directly benefit from ongoing advances in general-purpose image autoencoders.
To bridge the distributional gap between natural images and meteorological variables, we apply simple monotone transformations prior to encoding. This enables stable latent embedding and high-fidelity reconstruction even for highly skewed variables such as precipitation, as demonstrated in Supplementary Fig.~\ref{fig:s_recon_err_precip}.

From a practical perspective, diffusion-based DS remains computationally expensive at inference time compared to feed-forward alternatives, making improvements in sampling efficiency essential for real-world applications, including approaches discussed in \citep{hess2025fast}.
While diffusion models naturally enable ensemble generation for uncertainty-aware evaluation, recent work suggests that such ensembles may be under-dispersed relative to true errors \citep{mardani2025residual}, highlighting the need for more reliable uncertainty quantification in DS.

Our evaluation is limited to a single domain and a specific reanalysis-driven DS setting; assessing robustness across regions, forcing datasets, and future GCM projections remains important.
Although temporal leakage is mitigated using a strictly chronological, non-overlapping split, this setup does not fully capture nonstationarity or distribution shifts, and evaluation on truly future data is still required.

\section{Methods} \label{sec:methods}

\subsection{Datasets}

We use the Japanese 55-year Reanalysis (JRA-55) dataset \citep{kobayashi2015jra} as the low-resolution (LR) conditioning input and the Agro-Meteorological Grid Square Data (AMGSD) \citep{OHNO2016J-16-028} as the high-resolution (HR) reference.
Although reanalysis data differ from GCM projections, their physical consistency and temporal alignment with high-resolution observations enable a controlled hindcast-style evaluation of multivariate dependence diagnostics, making them a suitable testbed for assessing statistical DS performance.

The AMGSD fields are regridded to a $0.025^\circ \times 0.025^\circ$ latitude--longitude grid and cropped to the region 130.625--140.625$^\circ$E and 30.625--40.625$^\circ$N, yielding an HR spatial size of $400 \times 400$ \citep{oyama2023deep}.
Note that the HR reference data are provided only over land.
For JRA-55, the native horizontal resolution is $1.25^\circ \times 1.25^\circ$, which corresponds to an $8 \times 8$ grid over the same 10$^\circ \times$10$^\circ$ region; thus, the DS task corresponds to a 50$\times$ increase in linear resolution in each horizontal direction.
For JRA-55, daily values are derived by averaging the 3-hourly forecast outputs at 15Z, 18Z, 21Z, 00Z, 03Z, 06Z, 09Z, and 12Z to match the JST-based daily definition used in AMGSD. Daily maximum and minimum temperatures are computed from these 3-hourly values. For sea level pressure, daily means are computed by averaging the analysis fields at 18Z, 00Z, 06Z, and 12Z.

We use daily data for the period 1980--2023, with 1980--2000 for training, 2001--2003 for validation, and 2004--2023 for testing.
We downscale five HR target variables (Supplementary Table~\ref{tab:s_hr_variables}).
These targets cover temperature statistics, precipitation, and solar radiation, and include both strongly skewed (precipitation) and near-Gaussian (temperature) marginal distributions.
The LR conditioning inputs comprise nine meteorological variables (Supplementary Table~\ref{tab:s_lr_variables}) together with two static HR geographic fields (land--sea mask and topography) provided on the same $400 \times 400$ grid.

\subsection{Overview of the proposed diffusion-based DS method}

Our proposed method performs multivariate DS by modeling the joint distribution of the five HR target variables conditioned on LR reanalysis predictor fields and static geographic fields.
The core generative model is a conditional latent diffusion model (LDM) \citep{rombach2022high}, configured to be consistent with Stable Diffusion v1.5 (SD1.5) \citep{sd1.5}.
The diffusion formulation itself is unchanged; instead, we introduce several targeted adaptations to make the LDM applicable to meteorological DS, including (i) variable transformations that enable reuse of a pretrained autoencoder originally developed for natural images and (ii) conditioning of the reverse diffusion process on LR reanalysis predictor fields and static geographic fields; the details of these adaptations are described below.
In addition, we disable classifier-free guidance \citep{ho2021classifier} for the reasons discussed in the Discussion section, and we use the $v$-prediction formulation \citep{salimans2022progressive} instead of the $\epsilon$-prediction \citep{ho2020denoising} used in SD1.5, because the former achieved better DS performance in our preliminary experiments.

\begin{figure}[t]
    \centering
    \includegraphics[width=1.0\linewidth]{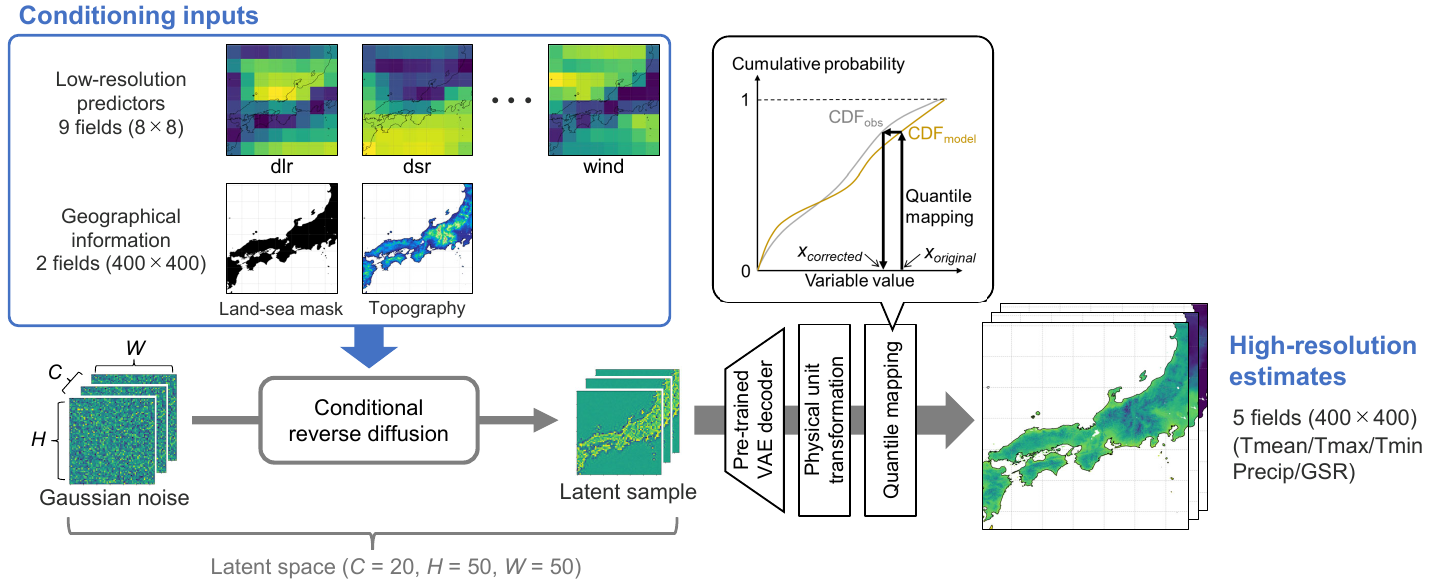}
    \caption{Low-resolution reanalysis predictors and geographic fields condition the latent diffusion model, which generates multivariate high-resolution meteorological fields. QM then corrects variable- and grid-point-wise biases in the generated outputs.}
    \label{fig:inference_overview}
\end{figure}

Overall, the training phase of the proposed method proceeds in three stages: (1) transform and encode the five HR target variables into a concatenated latent representation, (2) train a conditional diffusion model in this latent space, using the training data, conditioned on LR reanalysis predictor fields and static geographic fields, and (3) compute the empirical cumulative distribution functions (CDFs) of GT and the diffusion-based estimates, respectively, using the validation data, which are required for QM.
The inference phase of the proposed method comprises two conceptual stages (Fig.~\ref{fig:inference_overview}): diffusion-based multivariate generation followed by bias correction. A conditional latent diffusion model generates joint latent representations conditioned on coarse-resolution predictors, which are decoded and transformed back to physical space to obtain high-resolution meteorological fields. The resulting fields are then adjusted using quantile mapping (QM) to correct univariate distributions.

\subsection{Adapting a general image autoencoder for meteorology}

We reuse the pretrained autoencoder associated with the LDM to embed HR meteorological fields into a low-dimensional latent space.
Because meteorological variables have distributions that differ substantially from those of natural images, we apply simple monotone variable transformations prior to encoding.
This step is particularly important for precipitation, whose distribution is highly skewed and strongly concentrated near zero.
After transformation, each variable is embedded into a latent representation using the encoder, and the corresponding inverse transformation is applied to the decoder output to recover predictions in physical units.

Because the pretrained autoencoder expects three-channel RGB inputs, each meteorological variable---originally represented as a single channel---is replicated across three channels prior to encoding.
After decoding, the channel-wise mean is computed to recover a single-channel field.
All DS-target meteorological variables are embedded using the same pretrained autoencoder, and their resulting latent representations are concatenated along the channel dimension.
This concatenated latent representation serves as the target modeled by the diffusion process (see Supplementary Methods~\ref{sec:s_embedding_with_vae} for details).

\subsection{Conditioning on reanalysis predictors and geographic fields}

During training of the latent-space diffusion model, conditioning information is provided in the form of LR reanalysis predictor data corresponding to the same dates as the target HR fields, together with topographic data.
Note that the conditioning variables are not required to correspond one-to-one with the downscaled variables and may include additional auxiliary meteorological variables that are informative for the DS task.
This conditioning strategy is conceptually analogous to that used in LDMs, in which text prompts---embedded by a text encoder---are employed to condition the reverse diffusion process.
In this study, we employ both cross-attention and concatenation at the DNN input as conditioning mechanisms (see Supplementary Methods~\ref{sec:s_dnn_model} for details).

\subsection{Post-processing with QM}

QM is applied to the outputs generated by the trained diffusion model after decoding with the autoencoder and applying the inverse transformations.
Using the validation data, we perform diffusion-based DS and compute empirical CDFs for each variable at each HR grid point.
Empirical CDFs are likewise computed from the corresponding HR reference data.
For the test dataset, the diffusion-based DS results are corrected using variable- and grid-point-wise QM based on these empirical CDF pairs, and the bias-corrected fields are used as the final downscaled outputs (see Supplementary Methods~\ref{sec:s_bias_correction} for details).
Although we adopt QM in this study, the proposed framework is not limited to QM, and other bias-correction methods can be substituted if desired.

\subsection{Baselines}

We compare the proposed diffusion-based DS approach with three baseline families: interpolation with bias correction, a generative adversarial network (GAN)-based DNN super-resolution model, and diffusion-only generation without bias correction.
All methods are evaluated on the same data splits using the metrics described in the next subsection.

As a simple statistical baseline, we apply bilinear interpolation to upscale the LR fields to the HR grid and then perform QM at each variable and HR grid point using the training dataset (Interp-QM).
As the LR variable corresponding to GSR in the HR variables, we use surface downward shortwave radiation (Supplementary Table~\ref{tab:s_lr_variables}) after applying the unit conversion.

As a DNN-based baseline, we use $\pi$SRGAN \citep{oyama2023deep}, which extends a GAN-based DS framework \citep{stengel2020adversarial} by incorporating additional physically relevant input and reference variables.
In this study, we use all nine LR variables (Supplementary Table~\ref{tab:s_lr_variables}) as inputs to $\pi$SRGAN.
To achieve a magnification factor of 50, we adopt a two-stage cascade: a 10$\times$ DS model followed by a 5$\times$ DS model, trained independently for each stage, as in previous work \citep{stengel2020adversarial,oyama2023deep}.
We report results for both the raw $\pi$SRGAN outputs ($\pi$SRGAN) and their QM-corrected version ($\pi$SRGAN-QM), where QM is applied in the same manner as for the diffusion-based outputs.

To isolate the effect of bias correction, we also report diffusion-only DS results (Diffusion), in which the bias-correction step shown in Fig.~\ref{fig:inference_overview} is omitted, together with results obtained using QM (Diffusion-QM).

\subsection{Evaluation metrics}

We evaluate each model by comparing downscaled outputs for all test days with the corresponding HR reference fields.
Because the HR reference data are available only over land, all metrics are computed after masking out ocean grid cells.
Definitions of the evaluation metrics are provided in Supplementary Methods~\ref{sec:s_metrics_details}.

Because diffusion-based DS is stochastic, we perform inference with ten different random seeds.
For RMSE, precipitation distribution KLD, and SPEI3 evaluation, we use the ensemble-mean prediction obtained by averaging the ten samples.
For spatial and inter-variable correlation errors, we instead use a single sample (the first seed, $\mathrm{seed}=0$), as averaging across seeds smooths spatial patterns and weakens correlation structures, leading to degraded correlation-based scores.

\section*{Acknowledgements}
Numerical experiments were partially performed using the Earth Simulator at the Japan Agency for Marine-Earth Science and Technology under project number 1-25022.

\appendix
\renewcommand{\thesection}{S\arabic{section}}

\section{Supplementary Methods}

\subsection{Embedding meteorological variables with a pretrained VAE for natural images} \label{sec:s_embedding_with_vae}

Following LDM~\citep{rombach2022high}, instead of training a diffusion model that directly generates $400 \times 400$ HR fields, we first map the HR images to a lower-dimensional latent representation using a VAE~\citep{kingma2013auto} and then train the diffusion model in this latent space.

Although one could train a VAE specifically for meteorological variables, the cost of learning a VAE that can accurately reconstruct high-resolution images is generally high; accordingly, many techniques have been proposed to improve high-fidelity VAE reconstruction~\citep{higgins2017beta,gulrajani2017pixelvae,van2017neural,larsen2016autoencoding}.
Meanwhile, several pretrained VAE models trained on natural images are available as open-source resources.
We therefore reuse such an off-the-shelf pretrained VAE for meteorological data processing; in this study, we use the VAE from Kandinsky-3~\citep{kandinsky3}.
To apply a VAE trained on natural images to meteorological variables, we introduce the following adaptations.

\paragraph{Handling missing values in HR data}
The HR reference data used in this study are available only over land, and grid cells over the ocean are treated as missing values.
A straightforward approach to handle these missing regions is to assign a variable-specific constant (e.g., $-50$ for temperature) and treat it as a provisional ground-truth value in the missing regions (while excluding the missing regions when computing evaluation metrics).
However, constant imputation can introduce discontinuities near coastlines, which may lead to extremely large errors.
For example, when applying constant imputation to temperature, unrealistically low values close to $-50$ can be generated over land near the land--ocean boundary.

To avoid this issue, our method does not use constant values in the missing (ocean) regions.
Instead, we impute missing HR values using the value at the nearest land grid point.
This nearest-neighbor imputation mitigates abrupt changes across the land--ocean boundary and suppresses the extreme errors described above.

\paragraph{Handling differences in channel count}
Pretrained VAEs for natural images typically assume three-channel RGB inputs.
In contrast, each HR meteorological variable used in this study (Table~\ref{tab:s_hr_variables}) is provided as a single-channel field on a $400 \times 400$ grid.
We therefore replicate each variable across channels to form a three-channel input and then apply the VAE encoder to obtain the latent representation.

With the VAE encoder used in our experiments, an input of size $3 \times 400 \times 400$ (channel $\times$ height $\times$ width) is mapped to a latent tensor of size $4 \times 50 \times 50$.
We apply this encoding procedure independently to each of the five HR target variables in Table~\ref{tab:s_hr_variables} and concatenate the resulting latents along the channel dimension, yielding a $20 \times 50 \times 50$ latent tensor that is modeled by the diffusion process.

To map a generated latent tensor back to meteorological variables, we split the $20 \times 50 \times 50$ latent into five groups of four channels and apply the VAE decoder to each group, producing five reconstructed images of size $3 \times 400 \times 400$.
We then take the channel-wise mean of each reconstructed image to obtain a $1 \times 400 \times 400$ field, which serves as the final DS output for the corresponding variable.

\paragraph{Handling distribution mismatch}
Pixel intensities in RGB images take integer values between 0 and 255, and VAEs for natural images are typically trained on inputs normalized to the range $[0,1]$ by dividing by 255.
Accordingly, a VAE trained on natural images can be regarded as having learned an encoder--decoder pair that can map inputs whose values lie in $[0,1]$ to a latent space and back.

To match this assumption, we transform each HR variable before applying the VAE encoder, and apply the corresponding inverse transform after decoding to recover values in physical units.
Specifically, we use
\begin{align*}
  f_\mathrm{scale}(x) := \frac{x - x_\mathrm{min}}{x_\mathrm{max} - x_\mathrm{min}},
\end{align*}
where $x_\mathrm{min}$ and $x_\mathrm{max}$ are parameters chosen so that $x_\mathrm{min}$ and $x_\mathrm{max}$ in the original units are mapped to 0 and 1, respectively.
We estimate $x_\mathrm{min}$ and $x_\mathrm{max}$ from the distributions in the training data; the values used in this study are summarized in Table~\ref{tab:s_hr_variables}.
We apply the same transformation to the LR variables (Table~\ref{tab:s_lr_variables}).

\begin{figure}[htbp]
  \centering
  \includegraphics[width=\linewidth,clip]{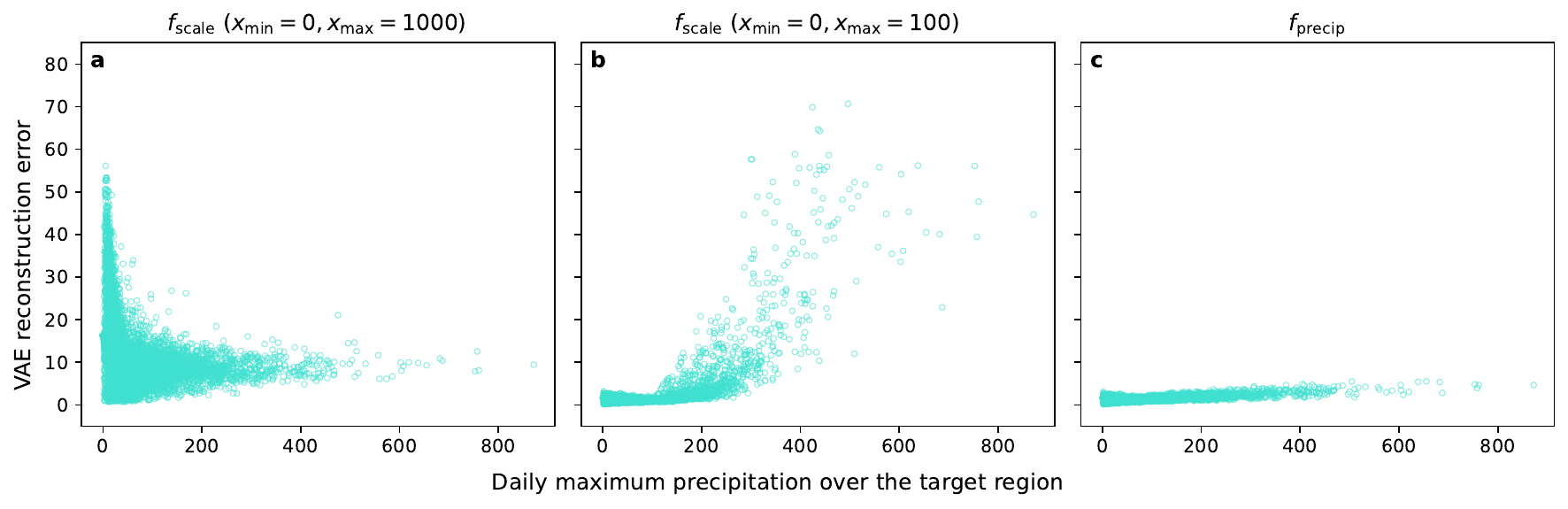}
  \caption{VAE reconstruction error for precipitation (vertical axis). The horizontal axis shows the daily maximum precipitation over the target region. Each point corresponds to one day in the training dataset. (a) Precipitation transformed with $f_\mathrm{scale}$ ($x_\mathrm{min}=0$, $x_\mathrm{max}=1000$) before VAE embedding. (b) Same as (a) but using $f_\mathrm{scale}$ with $x_\mathrm{min}=0$, $x_\mathrm{max}=100$. (c) Precipitation transformed with $f_\mathrm{precip}$ before VAE embedding.}
  \label{fig:s_recon_err_precip}
\end{figure}

Empirically, we found that this simple scaling enables VAE embedding and reconstruction for most meteorological variables.
However, for some variables---notably precipitation among the HR targets---the reconstruction quality remains poor when using $f_\mathrm{scale}$.
Precipitation is nonnegative and can reach values of roughly 900 in our dataset (mm\,day$^{-1}$).
A natural choice is to set $x_\mathrm{min}=0$ and $x_\mathrm{max}=1000$ for $f_\mathrm{scale}$; however, as shown in Fig.~\ref{fig:s_recon_err_precip}a, this leads to very large reconstruction errors.
Here, the reconstruction error is defined as the root mean squared error between $x$ and $x_\mathrm{recon}$, where $x_\mathrm{recon}$ is obtained by encoding $x$ to a latent variable $z$ and decoding $z$ back to the input space.
The horizontal axis in Fig.~\ref{fig:s_recon_err_precip} corresponds to the daily maximum precipitation over the target region.
The figure indicates that reconstruction errors are particularly large for samples with small precipitation values.
One possible reason is the difference in effective quantization near zero: dividing RGB pixel values by 255 yields increments of $1/255 \approx 0.0039$, whereas scaling precipitation by 1000 yields a finer increment of 0.001.
Such a mismatch may degrade reconstruction for samples concentrated near zero precipitation.

We next set $x_\mathrm{min}=0$ and $x_\mathrm{max}=100$ and apply $f_\mathrm{scale}$ (Fig.~\ref{fig:s_recon_err_precip}b).
In this case, reconstruction errors are reduced for precipitation below 100, whereas errors increase substantially for days with precipitation exceeding 100.
This is likely because the transformed values can exceed the VAE's assumed input range of $[0,1]$, leading to degraded reconstructions.

\paragraph{A nonlinear transformation for precipitation}
As shown in Fig.~\ref{fig:s_recon_err_precip}b, dividing precipitation by 100 before inputting it to the VAE reduces reconstruction errors when precipitation remains below 100.
Motivated by this observation, we propose the following transformation for precipitation:
\begin{align*}
  f_\mathrm{precip}(x) := \frac{\alpha}{100} \log \left(1 + \frac{x}{\alpha}\right) .
\end{align*}
The proposed $f_\mathrm{precip}$ satisfies $\left. f_\mathrm{precip}'(x) \right|_{x=0}=1/100$ regardless of $\alpha (>0)$, meaning that near $x=0$ it behaves equivalently to dividing by 100.
Moreover, $f_\mathrm{precip}$ is monotonically increasing, and its inverse is given by
\begin{align*}
  f_\mathrm{precip}^{-1}(y) = \alpha \left(\exp \left(\frac{100}{\alpha}y\right) - 1\right) .
\end{align*}
In this study, we set $\alpha=0.288 \times 10^2$ so that $f_\mathrm{precip}(900) \approx 1$.
Because it is difficult to obtain an analytic solution for $\alpha$ that exactly satisfies $f_\mathrm{precip}(900)=1$, we determine $\alpha$ numerically to approximately meet this condition.

Using $f_\mathrm{precip}$ enables a transformation that is close to dividing by 100 for small precipitation values while ensuring that even large precipitation values are mapped into the VAE's assumed input range of $[0,1]$.
To convert the VAE decoder output back to precipitation, we apply the inverse transform $f_\mathrm{precip}^{-1}$.
Figure~\ref{fig:s_recon_err_precip}c shows the reconstruction error obtained when transforming precipitation with $f_\mathrm{precip}$ and applying VAE reconstruction; the results confirm small reconstruction errors across a wide range of precipitation values.

\subsection{DNN model, conditioning, training, and inference} \label{sec:s_dnn_model}

\paragraph{DNN model}

As the DNN backbone for our diffusion model, we use the UNet architecture employed in Stable Diffusion~1.5~\cite{rombach2022high}\footnote{We use the configuration file at \url{https://huggingface.co/stable-diffusion-v1-5/stable-diffusion-v1-5/blob/main/unet/config.json} (accessed 2026-02-04), except that we set \texttt{block\_out\_channels} to \texttt{[160, 320, 640, 640]}.}.
This UNet is composed of convolutional layers~\cite{krizhevsky2012imagenet} and Transformer-based attention layers~\cite{NIPS2017_3f5ee243}.

\paragraph{Conditioning}
To train a diffusion model for DS task, we incorporate conditioning variables (LR reanalysis predictors and geographic fields) into the UNet using the following two mechanisms, which are used jointly.

We concatenate the nine LR reanalysis variables listed in Table~\ref{tab:s_lr_variables} along the channel dimension to form a $9 \times 8 \times 8$ tensor, and apply a convolutional projection to obtain a $64 \times 8 \times 8$ representation.
We then reshape it into a sequence of $8 \times 8$ tokens (each a 64-dimensional vector) and use it as the conditioning input to the cross-attention modules.

Following the approach used in diffusion-based super-resolution~\citep{rombach2022high}, we concatenate the conditioning variables with $x_t$ along the channel dimension, where $x_t$ is the latent variable modeled by our diffusion process (with shape $20 \times 50 \times 50$) and serves as the input to the DNN at timestep $t$ of the reverse diffusion process.
Specifically, we concatenate the nine LR reanalysis variables to form a $9 \times 8 \times 8$ tensor, upsample it to $9 \times 50 \times 50$ using interpolation, and concatenate it with $x_t$.
In addition, we concatenate the two geographic variables to form a $2 \times 400 \times 400$ tensor, downsample it to $2 \times 50 \times 50$, and concatenate it with $x_t$.

\paragraph{Training}
The diffusion model is trained using the AdamW optimizer~\cite{loshchilov2018decoupled} with a mini-batch size of 16 for 100 epochs. A weight decay of 0.01 is applied, and gradient clipping~\cite{10.5555/3042817.3043083} is used to stabilize training, with the gradient norm clipped to a maximum value of 1.  
Training is performed on a single NVIDIA A100 GPU, and the DNN model weights are cast to bfloat16 during training.

\paragraph{Inference}
During generation with the trained DNN model, we perform sampling using $T' = 50$ reverse diffusion steps. In diffusion-based image generation, it is common to use fewer time steps in the reverse process ($T' < T$) than those used during training. Following this practice, we set $T'$ to be smaller than the training time steps ($T = 1000$).

\subsection{Bias correction} \label{sec:s_bias_correction}
In the proposed framework, we apply a standard distribution-based bias-correction method to the HR estimates produced by the diffusion model.
Among the various bias-correction methods available, we adopt quantile mapping (QM) \citep{wood2004hydrologic,cannon2015bias}, one of the simplest and most widely used approaches.
Below, we describe the QM procedure used in this study.

Let $y \in \mathbb{R}^{n_v \times h \times w}$ denote the HR target fields, where $n_v$ is the number of variables and $h$ and $w$ are the spatial dimensions. The $(i,j)$-th component of variable $v$ is denoted by $y_{v,i,j} \in \mathbb{R}$.
Let $x$ denote the corresponding LR reanalysis predictor fields, and $\hat{y} \in \mathbb{R}^{n_v \times h \times w}$ the diffusion-model estimate of $y$ conditioned on $x$.
Let $\hat{F}_z$ denote the empirical cumulative distribution function (CDF) of a variable $z$ estimated from the dataset used for QM.

For non-learning-based methods such as Interp-QM, it is common to estimate the CDF from the training dataset and to perform bias correction based on that estimate.
In contrast, our diffusion model and $\pi$SRGAN are trained on the training dataset.
With sufficiently large model capacity and enough training steps, these learning-based models may overfit to the training data to some extent, and their predictive performance may therefore differ between the training data and holdout data, namely the validation and test datasets in our setting.
Because QM is intended to correct the holdout test data, it is more natural to estimate the CDF from holdout data of the same type.
Accordingly, in our experiments, for Diffusion and $\pi$SRGAN, we estimate the CDF used in QM from the validation dataset.
For the non-learning-based Interp-QM, we estimate the CDF from the training dataset.

For an LR input $x^\mathrm{test}$ in the test set, let $\hat{y}^\mathrm{test}$ denote the corresponding diffusion-model estimate.
QM defines, for each $(v,i,j)$, the bias-corrected estimate as
\begin{align*}
  \hat{y}_{v,i,j}^\mathrm{QM}
  = \hat{F}_{y_{v,i,j}}^{-1}
  \left(
    \hat{F}_{\hat{y}_{v,i,j}}
    \left(\hat{y}_{v,i,j}^\mathrm{test}\right)
  \right) .
\end{align*}
This operation maps $\hat{y}_{v,i,j}^\mathrm{test}$ to a value on the reference distribution $\hat{F}_{y_{v,i,j}}$ while preserving its quantile under the model-predicted distribution $\hat{F}_{\hat{y}_{v,i,j}}$, thereby correcting systematic distributional discrepancies between the model outputs and the HR reference data.
In practice, QM is applied separately for each calendar month, and we follow this standard approach in our experiments using the xsdba library~\citep{xsdba}.

\subsection{Evaluation metrics} \label{sec:s_metrics_details}
This section provides detailed definitions of the evaluation metrics used in the main text.
All metrics are computed on the test dataset after masking out ocean grid cells, since the HR reference data are available only over land.
For precipitation, we clip negative values in the downscaled outputs to zero before computing the metrics.

\paragraph{RMSE}
Let $\tilde{y}_{v,i,j}(t)$ denote the downscaled output for variable $v$ at HR grid point $(i,j)$ on day $t$, and let $y^{\mathrm{gt}}_{v,i,j}(t)$ denote the corresponding HR reference.
Let $\mathcal{I}_\mathrm{land}$ be the set of HR grid indices over land.
For each variable $v$ and day $t$, we compute
\begin{align}
  \mathrm{RMSE}_v(t)
  = \sqrt{
    \frac{1}{|\mathcal{I}_\mathrm{land}|}
    \sum_{(i,j)\in \mathcal{I}_\mathrm{land}}
    \left(
      \tilde{y}_{v,i,j}(t) - y^{\mathrm{gt}}_{v,i,j}(t)
    \right)^2
  } .
\end{align}
We report the temporal mean of $\mathrm{RMSE}_v(t)$ over the test period as the RMSE for variable $v$.

\paragraph{Inter-variable correlation error}
To assess preservation of multivariate dependence, we compute inter-variable correlations at each HR land grid point.
For each season (DJF, MAM, JJA, SON), Pearson correlation coefficients are computed over time between selected pairs of variables for both the downscaled outputs and the HR reference.
Let $\mathcal{P}$ denote the set of evaluated variable pairs, excluding combinations among Tmean, Tmax, and Tmin due to their strong deterministic dependence.
For each season, we compute the absolute error of the correlation coefficients for all pairs in $\mathcal{P}$ at each land grid point, and average over pairs and grid points.
We report these seasonal errors along with their mean across seasons (AVG).

\paragraph{Distribution KLD for precipitation}
Let $\mathcal{S}$ denote a set of representative sites (in our experiments, $|\mathcal{S}|=12$, corresponding to the sites used in \citep{oyama2023deep}).
For each site $s \in \mathcal{S}$, we construct empirical precipitation distributions over the test period from the downscaled and HR reference time series, and compute the Kullback--Leibler divergence (KLD) between them.
We report the mean KLD over $\mathcal{S}$ as the distribution KLD.

\paragraph{Spatial correlation error for precipitation}
We evaluate the reproducibility of precipitation spatial correlation patterns following \citep{oyama2023deep}.
For each representative site $s \in \mathcal{S}$, we compute Pearson correlation coefficients between the precipitation time series at $s$ and those at all other HR land grid points, yielding spatial correlation maps for both the downscaled output and the HR reference.
The mean squared error between these maps is then computed over land grid points and averaged across sites in $\mathcal{S}$ to obtain the Spatial correlation error.

\paragraph{IoU for severe drought area based on SPEI3}
We compute SPEI3 from the downscaled outputs and the HR reference using \citep{spei_R}, with potential evapotranspiration (PET) estimated following \citep{hargreaves1985reference}.
For each month, severe drought grid cells are defined as those with $\mathrm{SPEI3} \leq \delta$, where $\delta \in \{-1.0, -1.5, -2.0\}$ denotes the drought threshold. Spatial agreement between the downscaled and reference drought masks over land is then evaluated using the Intersection over Union (IoU), $\mathrm{IoU}(A,B) = |A \cap B| / |A \cup B|$ \citep{jaccard1901}.
We report the temporal mean of monthly IoU values, with higher values indicating better agreement in drought extent.

\section{Supplementary Tables and Figures}

\begin{table}[htbp]
  \centering
  \small
  \setlength{\tabcolsep}{4pt}
  \caption{HR variables and preprocessing transformations used in this study.}
  \label{tab:s_hr_variables}
  \begin{tabular}{l p{3.8cm} p{2.4cm} l p{4.0cm}}
    \toprule
    Variable & Description                     & Unit                     & Transform           & Parameters                   \\
    \midrule
    GSR      & Global solar radiation          & MJ\,m$^{-2}$\,day$^{-1}$ & $f_\mathrm{scale}$  & $x_{\min}=0,\ x_{\max}=60$   \\
    Precip   & Precipitation                   & mm\,day$^{-1}$           & $f_\mathrm{precip}$ & --                           \\
    Tmean    & 2\,m daily mean air temperature & $^{\circ}$C              & $f_\mathrm{scale}$  & $x_{\min}=-50,\ x_{\max}=50$ \\
    Tmax     & Daily maximum air temperature   & $^{\circ}$C              & $f_\mathrm{scale}$  & $x_{\min}=-50,\ x_{\max}=50$ \\
    Tmin     & Daily minimum air temperature   & $^{\circ}$C              & $f_\mathrm{scale}$  & $x_{\min}=-50,\ x_{\max}=50$ \\
    \bottomrule
  \end{tabular}
\end{table}

\begin{table}[htbp]
  \centering
  \small
  \setlength{\tabcolsep}{4pt}
  \caption{LR variables and preprocessing transformations used in this study.}
  \label{tab:s_lr_variables}
  \begin{tabular}{l p{3.8cm} p{2.4cm} l p{4.0cm}}
    \toprule
    Variable & Description                          & Unit           & Transform          & Parameters                    \\
    \midrule
    dlr      & Surface downward longwave radiation  & W\,m$^{-2}$    & $f_\mathrm{scale}$ & $x_{\min}=100,\ x_{\max}=450$ \\
    dsr      & Surface downward shortwave radiation & W\,m$^{-2}$    & $f_\mathrm{scale}$ & $x_{\min}=0,\ x_{\max}=450$   \\
    pr       & Precipitation                        & mm\,day$^{-1}$ & $f_\mathrm{scale}$ & $x_{\min}=0,\ x_{\max}=100$   \\
    rh2      & 2\,m relative humidity               & \%             & $f_\mathrm{scale}$ & $x_{\min}=0,\ x_{\max}=100$   \\
    slp      & Sea-level pressure                   & kPa            & $f_\mathrm{scale}$ & $x_{\min}=95,\ x_{\max}=105$  \\
    tmean    & 2\,m daily mean air temperature      & $^{\circ}$C    & $f_\mathrm{scale}$ & $x_{\min}=-50,\ x_{\max}=50$  \\
    tmax     & Daily maximum air temperature        & $^{\circ}$C    & $f_\mathrm{scale}$ & $x_{\min}=-50,\ x_{\max}=50$  \\
    tmin     & Daily minimum air temperature        & $^{\circ}$C    & $f_\mathrm{scale}$ & $x_{\min}=-50,\ x_{\max}=50$  \\
    wind     & Near-surface wind speed              & m\,s$^{-1}$    & $f_\mathrm{scale}$ & $x_{\min}=0,\ x_{\max}=20$    \\
    \bottomrule
  \end{tabular}
\end{table}

\begin{table}[htbp]
  \centering
  \small
  \setlength{\tabcolsep}{4pt}
  \caption{RMSE on the validation set for diffusion-model variants. For reference, results for Diffusion and Diffusion-QM are also shown. CFG denotes classifier-free guidance~\citep{ho2021classifier}, and $g$ denotes the guidance weight. The unconditional probability used for CFG during training is fixed to 0.1. ON denotes offset noise~\citep{offset_noise}, and $\sigma_{\mathrm{offset}}$ is its scale parameter.}
  \label{tab:s_val_rmse_cfg_offset}
  \begin{tabular}{l l c c c c c}
    \toprule
    Model         & Hyper-parameters               & Tmean & Tmax  & Tmin  & Precip & GSR   \\
    \midrule
    Diffusion-CFG & $g=1.2$                        & 1.722 & 2.748 & 1.713 & 5.812  & 3.936 \\
                  & $g=1.4$                        & 1.841 & 2.661 & 1.806 & 6.025  & 3.904 \\
                  & $g=1.6$                        & 1.979 & 2.627 & 1.926 & 6.268  & 3.877 \\
                  & $g=1.8$                        & 2.131 & 2.640 & 2.064 & 6.536  & 3.857 \\
                  & $g=2.0$                        & 2.293 & 2.697 & 2.215 & 6.828  & 3.844 \\
    \midrule
    Diffusion-ON  & $\sigma^2_\mathrm{offset}=0.1$ & 1.580 & 1.962 & 1.596 & 5.550  & 3.784 \\
                  & $\sigma^2_\mathrm{offset}=0.3$ & 1.374 & 1.831 & 1.816 & 5.824  & 2.630 \\
                  & $\sigma^2_\mathrm{offset}=0.5$ & 1.316 & 1.775 & 1.860 & 5.600  & 2.676 \\
    \midrule
    Diffusion     & --                             & 1.852 & 3.159 & 1.579 & 5.559  & 3.656 \\
    Diffusion-QM  & --                             & 0.890 & 1.268 & 1.238 & 5.501  & 2.325 \\
    \bottomrule
  \end{tabular}
\end{table}

\begin{figure}[htbp]
  \centering
  \includegraphics[width=0.7\linewidth]{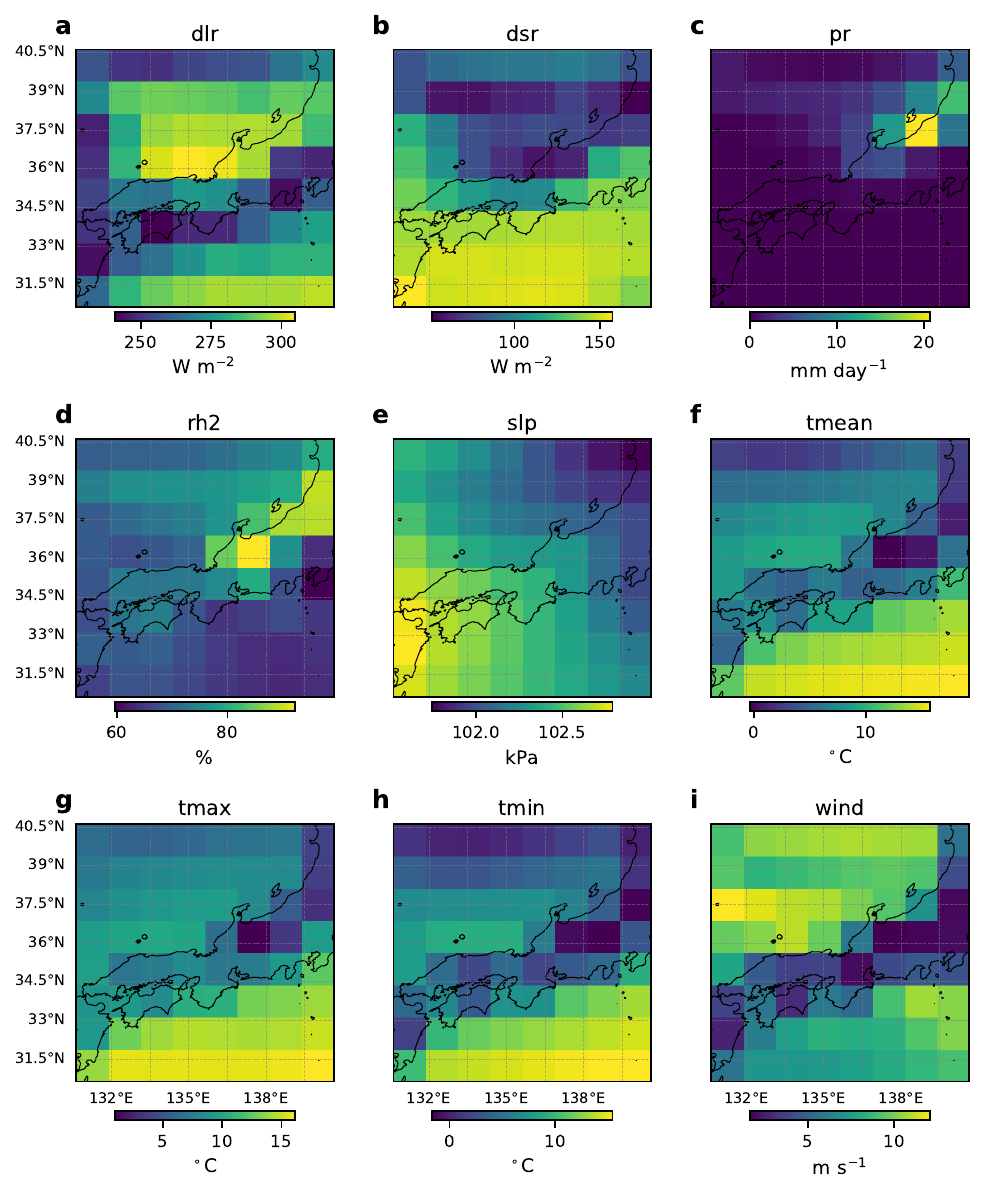}
  \caption{LR inputs for 2023-01-01, corresponding to the date used in Fig.~1 of the main manuscript. Panels show the reanalysis predictor fields used to condition the diffusion model.}
  \label{fig:s_lr_images_2023-01-01}
\end{figure}

\begin{figure}[htbp]
    \centering
    \includegraphics[width=1.0\linewidth,clip]{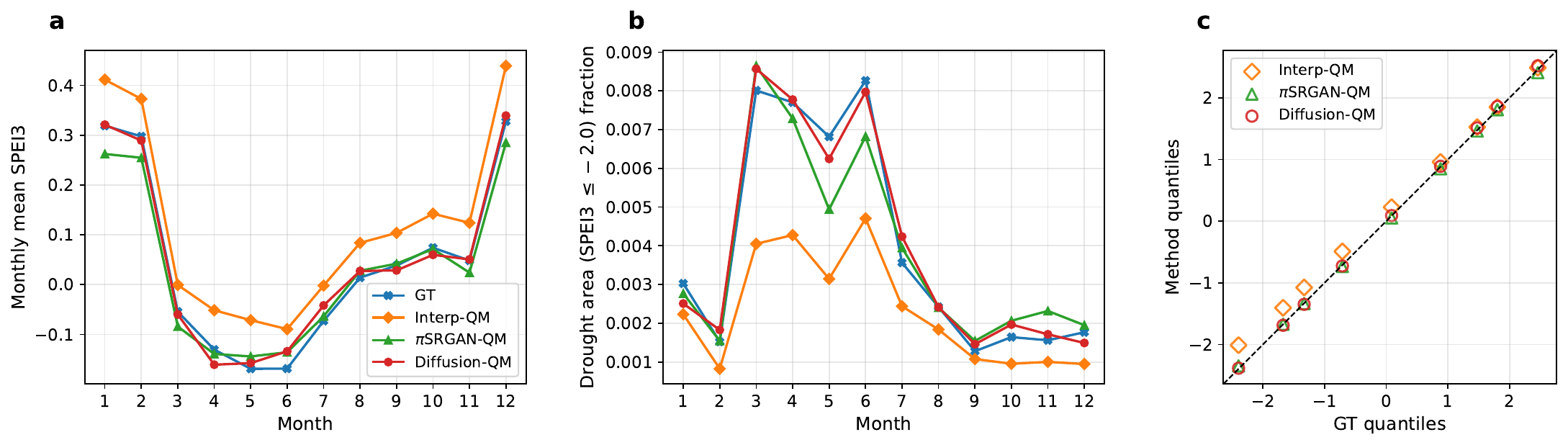}
    \caption{\textbf{a} Monthly mean SPEI3 over the test period, derived from GT and downscaled estimates. \textbf{b} Monthly fraction of the study area under severe drought conditions, defined by $\mathrm{SPEI3} \leq -2.0$. \textbf{c} Quantile--quantile comparison of SPEI3 values between GT and the downscaled estimates.}
    \label{fig:s_fig_monthly_spei_drought_qq}
\end{figure}

\bibliographystyle{naturemag}
\bibliography{../../references}

\end{document}